\documentclass[twocolumn,superscriptaddress,nofootinbib,amsmath,amssymb,aps,pra]{revtex4-1}

\usepackage{color}
\usepackage{braket}
\usepackage{graphicx}
\usepackage{dcolumn}
\usepackage{bm}
\usepackage{bbm}
\usepackage[mathlines]{lineno}

\newcommand{\note}[1] { {\color{red} [\uppercase{#1}] } }
\newcommand{\comment}[1] { {\color{blue} [\uppercase{#1}] } }
\newcommand{\ud}{\,\mathrm{d}}

\DeclareMathOperator{\tr}{tr}

\begin{document}

\title{Characterizing irreversibility in open quantum systems}

\author{Tiago B. Batalh{\~a}o}
\affiliation{Singapore University of Technology and Design, 8 Somapah Road, Singapore 487372}
\affiliation{Centre for Quantum Technologies, National University of Singapore, 3 Science Drive 2, Singapore 117543}
\author{Stefano Gherardini}
\affiliation{Department of Physics and Astronomy, LENS and QSTAR, University of Florence, via G. Sansone 1, I-50019 Sesto Fiorentino, Italy}
\author{Jader P. Santos}
\affiliation{Instituto de F\'isica da Universidade de Sa\~o Paulo, 05314-970 S\~ao Paulo, Brazil}
\author{Gabriel T. Landi}
\affiliation{Instituto de F\'isica da Universidade de Sa\~o Paulo, 05314-970 S\~ao Paulo, Brazil}
\author{Mauro Paternostro}
\affiliation{Centre for Theoretical Atomic, Molecular, and Optical Physics, School of Mathematics and Physics, Queen's University Belfast, BT7 1NN\\Belfast, United Kingdom}

\begin{abstract}
Irreversibility is a fundamental concept with important implications at many levels. It pinpoints the fundamental difference between the intrinsically reversible microscopic equations of motion and the unidirectional arrow of time that emerges at the macroscopic level. More pragmatically, a full quantification of the degree of irreversibility of a given process can help in the characterisation of the performance of thermo-machines operating at the quantum level. Here, we review the concept of entropy production, which is commonly intended as {\it the} measure of thermodynamic irreversibility of a process, pinpointing the features and shortcomings of its current formulation. 
\end{abstract}

\maketitle

When watching a movie, a question that can be made is whether the movie was recorded in that way, or if we are watching a time-reversed version of the actual recording. In our everyday experience, this question is often easy to answer, because watching broken pieces of glass moving from the floor to the top of a table and assembling themselves in the shape of a cup just feels weird. It is much more likely that the movie-makers recorded a glass cup falling down and breaking. Even though we are able to reach this conclusion quickly, none of the fundamental laws of physics (e.g., Newtonian equations of motion) forbid the broken pieces to reassemble the cup. Only the second law of thermodynamics makes the argument that, as the broken pieces represent a system of larger entropy, the reassembling process is impossible or at least very unlikely. 

The above is only a simple example, taken from everyday life, of a much deeper concept, namely that {\it the fundamental, microscopic equations of motion are symmetric under time-reversal, but the thermodynamical laws are not and establish a fundamental difference between past and future}. This apparent paradox has been known under the name of the ``arrow of time", given by Eddington in 1927 \cite{Eddington_ArrowOfTime}. The understanding of the emergence of the arrow of time from underlying quantum dynamics, and the formalisation of a self-consistent framework for its characterisation have been the focus of much interest. On one hand, we are in great need of tools able to reveal, experimentally, the implications that non-equilibrium dynamics has on the degree of reversibility for a given quantum process. On the other hand, the tools that are currently available for the (even only theoretical) investigation of thermodynamic irreversibility lack the widespread applicability and logical self-contained nature that is required from a complete theory. Let us elaborate more on this aspect. 

The entropy of an open system, unlike the energy, does not satisfy a continuity equation: in addition to  entropic fluxes exchanged between a system and its environment, some entropy may also be produced within the system. This contribution is called \emph{entropy production} and, according to the second law of thermodynamics, it is always non-negative, being zero only when the system and the environment are in thermal equilibrium. On the other hand, for closed systems that are dynamically brought out of equilibrium, the energy changes induced on a quantum system by a driving potential are not necessarily all translated into useful work that can be extracted from or performed on the system itself. Part of such energy is {\it lost} and gives rise to an entropic contribution akin to the entropy production above. 

Entropy production thus serves as {\it the} measure of the irreversibility of a physical process and may be used to characterize non-equilibrium systems in a broad range of situations  and across all scales. So far, several theories of entropy production have been developed in different contexts. Among the most significant examples in the classical domain are the theories formulated by Onsager~\cite{Onsager1931,Onsager1931a,Machlup1953,DeGroot1961,Tisza1957} and Schnakenberg~\cite{Germany1976,Tome2012}, which have been expanded upon towards their generalization to a wide range of classical stochastic processes~\cite{Tome2010,Spinney2012,Landi2013b,SeifertGabriel}.

The extensions of these approaches to small-scale (mesoscopic) systems made by Gallavotti, Cohen and collaborators~\cite{Evans1993,Evans1994,Gallavotti1995b}, Jarzynski~\cite{Jarzynski1997,*Jarzynski1997a}, and Crooks \cite{Crooks1998,Crooks2000}, and the more recent attempts at formulations that are fully within the quantum domain, have shown that quantum fluctuations may play a prominent role in determining the degree of irreversibility of non-equilibrium processes. 

This Chapter aims at addressing core questions in the formulation of entropy production and irreversibility in quantum systems and processes, both in the closed and open-system scenario~\cite{GooldChapter}. First, in Sec.~\ref{setting} we set the context and make founding statements on the relation between entropy production and irreversibility. We then move to Sec.~\ref{stochastic}, where we sketch a stochastic framework for entropy production. Finally, Sec.~\ref{alternative} is dedicated to the highlighting of conceptual shortcomings in the current formulation of entropy production, and the brief discussion of a potential alternative based on the use of a R\'enyi-2 entropy.

\section{The foundations of entropy production: closed-system dynamics}
\label{setting}

The goal of this Section is to address how the production of thermodynamic entropy is closely linked to a measure of distinguishability of past and future that can be cast as a guessing game and analysed with the tools of information theory and Bayesian reasonings. We will show that, by combining Bayes theorem from statistics with the Crooks fluctuation relation from non-equilibrium thermodynamics~\cite{Crooks2000}, it is possible to find an expression to quantify the level of certainty about the direction of the arrow of time \cite{Campisi_ArrowOfTime, Jarzinski_EqualitiesInequalities}.

We consider the Bayesian view of a guessing game, where the goal is to determine  the direction of the arrow of time of a given physical event (is a movie displayed in forward or time-reversed mode?), conditionally to a sequence of observations of the event itself. Let us call $F$ ($B$) the forward (backward) direction of the process. We assume that the \textit{a priori} probability for both directions is $1/2$.
{Bayes' theorem provides us  with the \textit{posterior} probability $P^F$ ($P^B$) that the event runs in the forward (backward) direction, on the basis of the the results of the observations, i.e. an explicit sequence of measurements. Explicitly
\begin{equation}
P^{F} = 1-P^B=\frac{1}{1+e^{-\Sigma}},
\label{eq:bayesposterior}
\end{equation}
where} $\Sigma$ is the (adimensional) entropy production associated with the system trajectory. When $\Sigma \gg 1$, then $P^{F} \approx 1$, and similarly $P^{F} \approx 0$ when $\Sigma \ll -1$. {Moreover, if $\Sigma = 0$ (no entropy production), then $P^{F} =P^{B} = 1/2$. The ratio between $P^{F}$ and $P^{B}$ is thus given by the ratio of the corresponding likelihood functions, which is expressed by the Crooks fluctuation theorem. Accordingly,} the ability to distinguish between the two directions of the arrow of time is directly linked to the entropy production along the observed trajectories.

Let us now assume that the event to witness is embodied by a process where the Hamiltonian of a quantum system experiences a change from time $t=0$ to $t=\tau$, thus generating an evolution of its initial state towards the state $\rho^{\text F}_t$. In doing so, an amount of thermodynamic work $W$ is done on or by the system, and the free energy of the system changes by $\Delta F$. The reverse process, then, would correspond to the driving of the state of the system towards the state $\rho^{\text B}_{\tau-t}$. Moreover, we assume that the initial state of the system is a thermal one at inverse temeperature $\beta$. For quantum systems, both thermal and quantum fluctuations conjure to determine the values taken by thermodynamically relevant quantities, including $W$, and the entropy production is written as $\Sigma=\beta(W-\Delta F)$. Its mean value can be shown to be given by the Kullback-Leibler relative entropy between the trajectories of time-opposite processes \cite{Kawai_PhaseSpace,Jarzinski_Lag,Kawai_ArrowOfTime}
\begin{equation}
\left\langle \Sigma \right\rangle = \beta \left\langle W \right\rangle - \beta \Delta F = {S}\left( \rho_t^{\text{F}} \Vert \rho_{\tau-t}^{\text{B}} \right),
\label{eq:entropyandkullback}
\end{equation}
where ${S}(\rho_a ||\rho_b)\equiv\text{Tr}[\rho_a(\ln\rho_a-\ln\rho_b)]$ for any pair of density matrices $\rho_{a,b}$.
In contrast with the macroscopic notion of distinguishability given by the guessing game, Eq.~(\ref{eq:entropyandkullback}) provides a microscopic measure of distinguishability of the dynamics between two processes in terms of their quantum representations in the Hilbert space, and is a nice link between the phenomenology of stochastic thermodynamics induced by a general quantum process and an information theoretical figure of merit for the difference between the states of the system in the forward and backward directions. 

Eqs.~(\ref{eq:bayesposterior}) and (\ref{eq:entropyandkullback}) were derived with some key assumptions, common to the Crooks's theorem, i.e. that the Hamiltonian depends explicitly on time and that the initial state of the system (in both forward and backward processes) corresponds to a Gibbs equilibrium. The first hypothesis breaks time homogeneity, and leads to the emergence of an arrow of time, while the second does not come from the microscopic equations of motion, as the dynamical equations that propagate the state of the system in time are not applied for $t < 0$. Instead, the state at $t=0$ is given and the dynamics is calculated only for $t>0$. This distinct treatment of past and future enforces the direction of the arrow of time.

It is worth observing that blurriness of the arrow of time does not depend explicitly on the size of system but occurs when the scale of energy changes in the system are comparable to the thermal energy $1/\beta$. In a truly macroscopic system, the usual energy and entropy scales are much larger than such threshold. This justifies the assumption that the arrow of time has a well-defined direction, which in turn implies that processes that lower entropy are extremely unlikely. 

The measurement of entropy production, or even its theoretical calculation, are tricky propositions \cite{Landi_EntropyProduction,Brunelli2016}, as it depends non-linearly on the density matrix and thus cannot be directly associated with a quantum mechanical observable. Some expressions have been derived in the case of relaxation \cite{Schlogl_Relaxation}, transport \cite{Lebowitz_Book}, and general processes in open and closed quantum systems \cite{Lutz_GeneralizedClausius,Lutz_EntropyProduction,GherardiniEntropy}. Experiments have been performed in systems like biomolecules \cite{Bustamante_Experiment2002,Bustamante_ExperimentRNA}, colloidal particles \cite{Blickle_ExperimentColloidal}, levitated nanoparticles \cite{Gieseler_ExperimentNanoparticle}, nuclear magnetic resonance \cite{Batalhao_2014}, optomechanical systems and cavity Bose-Einstein condensates \cite{Brunelli2016}. Some more examples of experiments up to 2013 can be found in \cite{Seifert_ReviewExperiment,Ciliberto_ReviewExperiment}. In many of these experiments, the thermal energy is much higher than the separation of quantum energy levels, thus indicating that these experiments probed classical non-equilibrium thermodynamics. In the case of the NMR experiment reported in Ref.~\cite{Batalhao_2014}, instead, it was possible to observe quantum coherences between the energy eigenstates that were not wiped out by any decoherence process. Combined with the fact that the energy level separation was of the same order as the thermal energy, one can say that this is one example of an experiment in quantum thermodynamics. 

\section{Stochastic quantum entropy production}
\label{stochastic}


The definition of stochastic quantum entropy production $\sigma$ for an arbitrary open quantum system $\mathcal{S}$, and the characterization of its statistics, pass through the evaluation of the quantum fluctuation theorem for such system. As discussed in the preceding Section, the latter relies on establishing forward and backward protocols for a given non-equilibrium process, which define the difference between performing such a transformation in a direction or in its time-reversed version along the arrow of time. 


In small systems, negative entropy productions can occur during individual processes. Fluctuation theorems from stochastic thermodynamics can quantify the occurrence of such events, and, thus, the characterization of the statistics of $\sigma$ is crucial for determining irreversibility~\cite{GherardiniEntropy,GherardiniThesis}. If we want to measure the statistics of the entropy production of an arbitrary quantum system for each input and output measurement result, we can adopt the two-time quantum measurement scheme discussed in previous Chapters, that has to be in agreement with the fluctuation theorem \cite{Crooks2000,Campisi_ArrowOfTime,Jarzinski_EqualitiesInequalities}

To this aim, let us consider an open quantum system that undergoes a forward transformation in the interval $[0,\tau]$ consisting of measurement, dynamical evolution and second measurement. At time $t=0^-$ the system is prepared in a state $\rho_{0}$ and then subjected to a measurement of the observable $\mathcal{O}_{\textrm{in}}$, which is defined by the set $\{\Pi^{\textrm{in}}_{m}\}$ of projector operators given in terms of the $m^{\text{th}}$ possible outcomes of the first measurement of the protocol. After the first measurement at $t = 0^{+}$, the system undergoes a time evolution, which we assume to be described by a \textit{unital} completely positive, trace-preserving (CPTP) map $\Phi:L(\mathcal{H})\rightarrow L(\mathcal{H})$, with $L(\mathcal{H})$ denoting the sets of density operators defined on the Hilbert space $\mathcal{H}$. A CPTP map is unital if it preserves the identity operator $\mathbbm{1}$ on $\mathcal{H}$, {\em i.e.} $\Phi(\mathbbm{1}) = \mathbbm{1}$. The request of unitality covers a large family of quantum physical transformations not increasing the purity of the initial states, including, among others, unitary evolutions and decoherence processes. Such assumption, moreover, does not limit the generality of the approach, if the open quantum system $\mathcal{S}$ is considered to be a multipartite system~\cite{GherardiniEntropy,GherardiniThesis}. 

The time-evolved dynamics of the system is then denoted as $\rho_{\textrm{fin}} \equiv \Phi(\rho_{\textrm{in}})$; in case of unitary evolution with Hamiltonian $H(t)$, the final quantum state at $t = \tau^{-}$ is $\rho_{\textrm{fin}} = \Phi(\rho_{\textrm{in}}) = \mathcal{U}\rho_{\textrm{in}}\mathcal{U}^{\dagger}$, where $\mathcal{U}=\mathbb{T}\exp\left(-\frac{i}{\hbar}\int^{\tau}_{0}H(t)dt\right)$ is the unitary time evolution operator ($\mathbb{T}$ is the time-ordering operator). After the evolution at time $t = \tau^+$, the second measurement of the protocol is performed on the quantum system and the observable $\mathcal{O}_{\textrm{fin}}$ is measured, where $\Pi^{\textrm{fin}}_{k}$ is the projector operator related to the $k^{\text{th}}$ outcome $a^{\textrm{fin}}_{k}$. We denote with $\rho_{\tau}$ the resulting density operator, describing the ensemble average of the post-measurement state after the second measurement. For the forward process, in order to characterize the stochastic quantum entropy production we have to record only the joint probability $p(a^{\textrm{fin}}_{k}, a^{\textrm{in}}_{m})$ that the events ``measurement of $a^{\textrm{in}}_{m}$'' and ``measurement of $a^{\textrm{fin}}_{k}$'' both occur in a single realization, i.e.
\begin{equation}
p(a^{\textrm{fin}}_{k},a^{\textrm{in}}_{m}) = \textrm{Tr}\left[\Pi^{\textrm{fin}}_{k}\Phi(\Pi^{\textrm{in}}_{m}\rho_{0}\Pi^{\textrm{in}}_{m})\right].
\end{equation}
To derive the backward process $B$, it is essential to introduce the concept of time-reversal, which relies on the time reversal symmetry (or T-symmetry). A time reversal transformation $T_{R}$ overturns the time axis, i.e. $T_{R}:t\mapsto -t$. Let us stress again that time-symmetry is broken, in general, unless the system is in an equilibrium state. Time-reversal is achieved by a time-reversal operator $\Theta$, which acts on the system Hilbert space and has to be an antiunitary operator, since a symmetry operation on a quantum-mechanical system can be performed only by a unitary or antiunitary operator. An operator is antiunitary if 
\begin{itemize}
\item[(i)] It is anti-linear, i.e. $\Theta (x_1|\varphi_1\rangle + x_2|\varphi_2\rangle)= x_1^\star \Theta|\varphi_1\rangle + x_2^\star \Theta|\varphi_2\rangle$ for arbitrary complex coefficients $x_1$, $x_2$ and $|\varphi_1\rangle$, $|\varphi_2\rangle$ $\in$ $\mathcal{H}$; 
\item[(ii)] It transforms the inner product as $\langle \widetilde{\varphi}_1|\widetilde{\varphi}_2\rangle=\langle\varphi_2|\varphi_1\rangle$
for $|\widetilde{\varphi}_1\rangle=\Theta|\varphi_1\rangle$ and $|\widetilde{\varphi}_2\rangle=\Theta|\varphi_2\rangle$,  
\item[(iii)] It satisfies the relations $\Theta^{\dagger}\Theta = \Theta\Theta^{\dagger} = \mathbbm{1}$.
\end{itemize}
The fulfilment of each of these features ensures that $\Theta$ obeys the T-symmetry \cite{Sozzi2008Book}. Accordingly, we define the time-reversed density operator as $\widetilde{\rho}\equiv\Theta\rho\Theta^\dagger$. Then, in order to obtain the backward process $B$ we also need to introduce the time-reversal version of the quantum evolution of the system. A significant result, first shown in Ref.~\cite{CrooksPRA2008} and recently generalized in~\cite{Manzano2015}, states that the time-reversed quantum map $\widetilde{\Phi}$ of the CPTP map $\Phi$ is equally CPTP and admits an operator-sum (or Kraus) representation, obeying the relation $\sum_{u}\widetilde{E}_{u}^{\dagger}\widetilde{E}_{u} = \mathbbm{1}$. Accordingly, it shall be written as $\widetilde{\Phi}(\rho) = \sum_{u}\widetilde{E}_{u}\rho\widetilde{E}_{u}^{\dagger}$, where $\widetilde{E}_{u}$ is generally expressed as a function of $E^{\dagger}_{u}$ and  the invertible fixed point of the quantum map (notice that this might not necessarily be unique). In particular, for a unital CPTP quantum map $\widetilde{E}_{u} = \Theta E^{\dagger}_{u}\Theta^\dagger$. Now, we are in the position to define the backward process. At $t=\tau^+$ the system is prepared in the state $\widetilde{\rho}_{\tau} = \Theta\rho_{\tau}\Theta^\dagger$, and we measure the observable $\widetilde{\mathcal{O}}_{\textrm{ref}}$, that is defined by the projectors $\widetilde{\Pi}^{\textrm{ref}}_{k} = |\widetilde{\phi}_{a_{k}}\rangle\langle\widetilde{\phi}_{a_{k}}|$, with $|\widetilde{\phi}_{a_{k}}\rangle \equiv \Theta|\phi_{a_{k}}\rangle$. The first measurement of the backward process is chosen equal to the time-reversed version of the second measurement of the forward process, where the state after the first measurement of the backward process is usually called \textit{reference state}. In this regard, it is worth noting that although the quantum fluctuation theorem can be derived without imposing a specific operator for the reference state~\cite{Sagawa2014}, we have chosen that the reference state is identically equal to the final density operator after the second measurement of the forward process. This choice appears to be the most natural among the possible ones to design a suitable scheme for the measurement of the stochastic entropy production, consistently with the quantum fluctuation theorem and the asymmetry of the second law of thermodynamics. Afterwards, in the reversal direction of the arrow of time, the reference state undergoes the time-reversal dynamical evolution, mapping it onto the initial state of the backward process $\widetilde{\rho}_{\textrm{in}'} = \widetilde{\Phi}(\widetilde{\rho}_{\textrm{ref}})$, and at $t = 0^{+}$ the density operator $\widetilde{\rho}_{\textrm{in}'}$ 
is subject to the second projective measurement of the backward process, whose observable is given by $\widetilde{\mathcal{O}}_{\textrm{in}}$ and is defined by the projectors $\widetilde{\Pi}^{\textrm{in}}_{m} = |\widetilde{\psi}_{a_{m}}\rangle\langle\widetilde{\psi}_{a_{m}}|$ with $|\widetilde{\psi}_{a_{m}}\rangle\equiv\Theta|\psi_{a_{m}}\rangle$. As for the forward process, we compute the joint probability $p(a^{\textrm{in}}_{m}, a^{\textrm{ref}}_k)$ 
to simultaneously measure the outcomes $a^{\textrm{in}}_{m}$ and $a^{\textrm{ref}}_k$ in a single realization of the backward process
\begin{equation}
p(a^{\textrm{in}}_{m}, a^{\textrm{ref}}_{k}) = \textrm{Tr}[\widetilde{\Pi}^{\textrm{in}}_{m}\widetilde{\Phi}(\widetilde{\Pi}^{\textrm{ref}}_{k}\widetilde{\rho}_{\tau}
\widetilde{\Pi}^{\textrm{ref}}_{k})].
\end{equation}
As shown in Ref. \cite{GherardiniEntropy}, the combination of the two-time quantum measurement scheme with the quantum fluctuation theorem requires to perform the $2$nd and $1$st measurement of the backward protocol, respectively, on the same basis of the $1$st and $2$nd measurement of the forward process after the time-reversal transformation.

The following scheme well summarizes the forward and backward processes regarding the quantum fluctuation theorem:
\begin{eqnarray*}
\text{FORWARD}:~\rho_{0}\underbrace{\longmapsto}_{\{\Pi^{\textrm{in}}_{m}\}}\rho_{\textrm{in}}\underbrace{\longmapsto}_{\Phi}\rho_{\textrm{fin}}
\underbrace{\longmapsto}_{\{\Pi^{\textrm{fin}}_{k}\}}\rho_{\tau} \\
\text{BACKWARD}:~\widetilde{\rho}_{\tau}\underbrace{\longmapsto}_{\{\widetilde{\Pi}^{\textrm{ref}}_{k}\}}\widetilde{\rho}_{\textrm{ref}}
\underbrace{\longmapsto}_{\widetilde{\Phi}}\widetilde{\rho}_{\textrm{in}'}\underbrace{\longmapsto}_{\{\widetilde{\Pi}^{\textrm{in}}_{m}\}}\widetilde{\rho}_{0'}
\end{eqnarray*}
Now, we can define the stochastic quantum entropy production $\sigma$
\begin{equation}\label{general_sigma}
\sigma(a^{\textrm{fin}}_{k},a^{\textrm{in}}_{m}) \equiv \ln\left[\frac{p(a^{\textrm{fin}}_{k}, a^{\textrm{in}}_{m})}{p(a^{\textrm{in}}_{m}, a^{\textrm{ref}}_{k})}\right]
= \ln\left[\frac{p(a^{\textrm{fin}}_{k}|a^{\textrm{in}}_{m})p(a^{\textrm{in}}_{m})}
{p(a^{\textrm{in}}_{m}|a^{\textrm{ref}}_{k})p(a^{\textrm{ref}}_{k})}\right],
\end{equation}
thus providing a general expression of the quantum fluctuation theorem for the considered open quantum system subject to a two-time quantum measurement scheme. In Eq. (\ref{general_sigma}) $p(a^{\textrm{fin}}_{k}|a^{\textrm{in}}_{m})$ and $p(a^{\textrm{in}}_{m}|a^{\textrm{ref}}_{k})$ are the conditional probabilities of measuring, respectively, the outcomes $a^{\textrm{fin}}_{k}$ and $a^{\textrm{in}}_{m}$, conditioned on having first measured $a^{\textrm{in}}_{m}$ and $a^{\textrm{ref}}_{k}$. Its mean value
\begin{equation}\label{mean_sigma}
\langle\sigma\rangle = \sum_{k,m}p(a_k^{\textrm{fin}},a_m^{\textrm{in}})\ln\left[\frac{p(a_k^{\textrm{fin}},a_m^{\textrm{in}})}{p(a_k^{\textrm{in}}, a_m^{\textrm{ref}})}\right]
\end{equation}
corresponds to the classical relative entropy (or Kullback-Leibler divergence) between the joint probabilities $p(a^{\textrm{fin}},a^{\textrm{in}})$ and $p(a^{\textrm{in}},a^{\textrm{ref}})$ of the forward and backward processes, respectively. The Kullback-Leibler divergence is always non-negative~\cite{Cover2006}, and, thus, $\langle\sigma\rangle\geq 0$. Here, it is worth noting how these relations are strongly connected to the results highlighted in the previous Section of this Chapter. In particular, $\langle\sigma\rangle$ is effectively the amount of additional information that is required to achieve the backward process, once the quantum system has reached the final state $\rho_{\tau}$, and $\langle\sigma\rangle = 0$ if and only if $p(a^{\textrm{fin}}_{k},a^{\textrm{in}}_{m}) = p(a^{\textrm{in}}_{m},a^{\textrm{ref}}_{k})$, i.e. if and only if $\sigma = 0$.Thus, the transformation from $t=0^{-}$ to $t=\tau^{+}$ can be defined to be thermodynamically irreversible if $\langle\sigma\rangle > 0$. When, instead, all the fluctuations of $\sigma$ shrink around $\langle\sigma\rangle \simeq 0$ the system comes closer and closer to be reversible. We observe that a system transformation may be thermodynamically irreversible also if the system undergoes unitary evolutions with the corresponding irreversibility contributions due to applied quantum measurements. Also the measurements back-actions, indeed, lead to energy fluctuations of the quantum system. In case there is no evolution (identity map) and the two measurement operators are the same, then the transformation becomes reversible. Finally, we note that if the CPTP quantum map $\Phi$ is unital, then $p(a^{\textrm{fin}}_{k}|a^{\textrm{in}}_{m}) = p(a^{\textrm{in}}_{m}|a^{\textrm{ref}}_{k})$, and the stochastic quantum entropy production $\sigma$ turns out to be equal to
\begin{equation}\label{sigma}
\sigma(a^{\textrm{fin}}_{k},a^{\textrm{in}}_{m}) = \ln\left[\frac{p(a^{\textrm{in}}_{m})}{p(a^{\textrm{ref}}_{k})}\right] =\ln\left[\frac{\langle\psi_{a_{m}}|\rho_{0}|\psi_{a_{m}}\rangle}
{\langle\widetilde{\phi}_{a_{k}}|\widetilde{\rho}_{\tau}|\widetilde{\phi}_{a_{k}}\rangle}\right].
\end{equation}

\subsection{Connection with the system quantum relative entropy}

The irreversibility of an arbitrary system transformation within a two-time measurement scheme for an open quantum system in interaction with the environment is encoded in the mean stochastic entropy production $\langle\sigma\rangle$. In Ref. \cite{GherardiniEntropy}, it has been proved the relation between $\langle\sigma\rangle$ and the quantum relative entropy of the system density matrix in correspondence of the final time of the system transformation for unital CPTP quantum maps. Accordingly, under the hypotheses that (i) the reference state of the quantum fluctuation theorem is identically equal to the final density operator after the second measurement of the forward process and (ii) the first measurement of the backward process is chosen equal to the time-reversed version of the second measurement of the forward process (both hypotheses originate from the introduction of the two-time measurement scheme for the measure of $\sigma$), the quantum relative entropy $S(\rho_{\textrm{fin}}\parallel\rho_{\tau})$ fulfills the following inequality:
\begin{equation}\label{eq:entropy-positivity}
0\leq S(\rho_{\textrm{fin}}\parallel\rho_{\tau})\leq\langle\sigma\rangle,
\end{equation}
where the equality $S(\rho_{\textrm{fin}}\parallel\rho_{\tau}) = 0$ holds if and only if $\rho_{\textrm{fin}} = \rho_{\tau}$. Moreover, for $[\mathcal{O}_{\textrm{fin}},\rho_{\textrm{fin}}]=0$ one has $\langle\sigma\rangle = S(\rho_{\tau}) - S(\rho_{\textrm{in}})$, so that
\begin{equation}\label{eq:entropy-of-map}
0 = S(\rho_{\textrm{fin}}\parallel\rho_{\tau})\leq\langle\sigma\rangle=S(\rho_{\textrm{fin}})-S(\rho_{\textrm{in}}),
\end{equation}
where $S(\rho)= - \tr(\rho \ln \rho)$ is the von Neumann entropy. For a closed quantum system following a unitary evolution, $S(\rho_{\textrm{fin}}\parallel\rho_{\tau}) = \langle\sigma\rangle$. This result is in agreement with Eq.~(\ref{eq:entropyandkullback}) \cite{Kawai2007PRL98,Vaikuntanathan2009EPL87,Parrondo2009NJP11}, which provides a microscopic measure (in terms of the Kullback-Leibler relative entropy) of the distinguishability between the system trajectories, respectively, in the forward and backward process, when the initial state of the quantum system in both processes corresponds to a Gibbs equilibrium induced by an external thermal bath. While Eq.~(\ref{eq:entropy-positivity}) is more general and includes the irreversibility contributions of both the map $\Phi$ and the final measurement, in Eq.~(\ref{eq:entropy-of-map}) due to a special choice of the observable of the second measurement we obtain $\rho_{\textrm{fin}}=\rho_{\tau}$ and, thus, the quantum relative entropy vanishes while the stochastic quantum entropy production contains the irreversibility contribution only from the map, which is given by the difference between the von Neumann entropy of the final state $S(\rho_{\textrm{fin}})$ and the initial one $S(\rho_{\textrm{in}})$.

\subsection{Quantum entropy production statistics}

The statistics of the stochastic quantum entropy production $\sigma$ can be computed by evaluating the corresponding probability distribution $\textrm{Prob}(\sigma)$. Indeed, depending on the values assumed by the measurement outcomes $\{a^{\textrm{in}}\}$ and $\{a^{\textrm{fin}}\}$, $\sigma$ is a fluctuating variable. Thus, each time we repeat the experiment described above for the measure of the stochastic quantum entropy production, we shall have a different realization for $\sigma$, within a set of discrete values in case $\mathcal{S}$ is finite dimensional. The probability distribution of $\sigma$ is thus fully determined by the knowledge of the measurement outcomes and the respective probabilities (relative frequencies). In particular, let us consider again $p(a^{\text{fin}}_k,a^{\text{in}}_m)$, which denotes the joint probability to obtain the measurement outcomes $a^{\textrm{in}}_{m}$ and $a^{\textrm{fin}}_{k}$. We have 
\begin{equation}\label{joint_prob}
p(a^{\text{fin}}_k,a^{\text{in}}_m) = \textrm{Tr}\left[\Pi^{\textrm{fin}}_{k}\Phi(\Pi^{\textrm{in}}_{m})\right]p(a_{m}^{\textrm{in}}),
\end{equation}
where $p(a_{m}^{\textrm{in}})$ is the probability to obtain the measurement outcome $a_{m}^{\textrm{in}}$ after the first measurement of the forward process. Accordingly, the probability distribution $\textrm{Prob}(\sigma)$ turns out to be
\begin{eqnarray}\label{prob_sigma}
\textrm{Prob}(\sigma) &=& \left\langle\delta\left[\sigma - \sigma(a^{\textrm{in}}_{m},a^{\textrm{fin}}_{k})\right]\right\rangle\nonumber \\
&=& \sum_{k,m}\delta\left[\sigma - \sigma(a^{\textrm{in}}_{m},a^{\textrm{fin}}_{k})\right]p(a^{\text{fin}}_k,a^{\text{in}}_m),
\end{eqnarray}
where $\delta[\cdot]$ is the Dirac-delta distribution.

In the frequency domain, the properties of the corresponding probability distribution $\textrm{Prob}(\sigma)$ are completely defined by its Fourier transform, i.e. its characteristic function. The latter, similarly to the what is commonly done for the work and heat distribution~\cite{GherardiniBuffoni}, is a key quantity to be indirectly measured for the inference of the statistics of $\sigma$, and, thus, of the irreversibility for an open quantum system. The characteristic function $G(u)$ of the probability distribution $\textrm{Prob}(\sigma)$, with $u\in\mathbb{C}$ complex number, is defined as
\begin{equation}\label{G}
G(u) = \int \textrm{Prob}(\sigma)e^{iu\sigma} d\sigma.
\end{equation}
By substituting in Eq. (\ref{G}) the expression of the probability distribution $\textrm{Prob}(\sigma)$ and exploiting the linearity of the CPTP quantum maps and of the trace (with $\Phi$ unital), the characteristic functions can be written in the following form:
\begin{eqnarray}
G(u) = \mathrm{Tr}\left[\rho_{\tau}^{-iu}\Phi(\rho_{\mathrm{in}}^{1+iu})\right]\,,
\end{eqnarray}
that will be used to effectively measure the thermodynamic irreversibility of $\mathcal{S}$. Moreover, by choosing $u = i$, we recover the Jarzynski identity 
$\big\langle e^{-\sigma}\big\rangle \equiv G(i) = \textrm{Tr}\left\{\rho_{\tau}\Phi\left[\mathbbm{1}\right]\right\} = 1$
for the stochastic quantum entropy production $\sigma$, also called \textit{integral quantum fluctuation theorem} \cite{Sagawa2014}.

\subsection{Measuring irreversibility}

Until now experiments for the measurement of the quantum entropy production statistics of an open quantum system have not been performed, even if in \cite{GherardiniEntropy} a procedure has recently been proposed. The latter is based on quantum estimation methods, and relies on the indirect measurement of the characteristic function $G(u = i\gamma) \equiv \big\langle e^{-\gamma\sigma}\big\rangle$ for a set of values of $\gamma\in\mathbb{R}$. In particular, the characteristic function $G(u = i\gamma)$ depends exclusively on suitable powers of the initial and final density operators of the quantum system $\mathcal{S}$, and these density operators are diagonal in the basis of the observable eigenvectors. Thus, they can be measured by means of standard state population measurements for each value of $\gamma$. This result can lead to a significant reduction of the number of measurements that is required to reconstruct the probability distribution $\textrm{Prob}(\sigma)$, beyond the direct application of the definition of Eq.~(\ref{prob_sigma}).

Here, we will show in a nutshell the procedure for the indirect measurement of the characteristic function $G(u = i\gamma)$, leading then to the statistics of $\sigma$: (i) Prepare the initial product state $\rho_{\textrm{in}} = \sum_{m}\Pi^{\rm fin}_{m}p(a^{\textrm{in}}_{m})$, which is diagonal in the basis composed by the eigenvectors of the first measurement observable $\mathcal{O}_{\rm in}$ and, thus, defined by the probabilities $p(a^{\textrm{in}}_{m})$. Then, after the quantum system is evolved within the time interval $[0,\tau]$, measure the occupation probabilities $p(a^{\textrm{fin}}_{k})$ and compute the stochastic quantum entropy production $\sigma$ as given in Eq. (\ref{mean_sigma}). (ii) For every chosen value of $\gamma$ (one possible choice for the optimal values of the set of real parameters $\{\gamma\}$ has been discussed in Ref. \cite{GherardiniEntropy,GherardiniThesis}), prepare the quantum system in the states $\hat{\rho}_{\textrm{in}}(\gamma) \equiv \rho_{\mathrm{in}}^{1 - \gamma}/{\rm Tr}[\rho_{\mathrm{in}}^{1 - \gamma}]$, and let it evolve. (iii) As we have that 
\begin{equation}
\begin{aligned}
G(i\gamma) &= \sum_{k}\sum_{m}\langle k|p(a^{\textrm{fin}}_{k})^{\gamma}|m\rangle\langle m|
\Phi\left(\hat{\rho}_{\textrm{in}}(\gamma)\right)|k\rangle \\
&= \sum_{k}p(a^{\textrm{fin}}_{k})^{\gamma}\langle k|\Phi\left(\hat{\rho}_{\textrm{in}}(\gamma)\right)|k\rangle,
\end{aligned}
\end{equation}
after performing a trace operation with respect to the orthonormal basis which spans the Hilbert space of $\mathcal{S}$,
measure the occupation probabilities $\langle k|\Phi\left(\hat{\rho}_{\textrm{in}}(\gamma)\right)|k\rangle$, so as to finally obtain $G(i\gamma)$.

It is worth observing that the measure of the characteristic functions $G(i\gamma)$ relies only on the measure of occupation probabilities, and, thus, the proposed procedure does not require full tomography. Moreover, quite remarkably, for the three steps of the procedure the required number of measurements scales linearly with the number of possible measurement outcomes (coming from the system at the initial and final stages of the transformation), or equivalently with the number of values that can be assumed by the stochastic quantum entropy production $\sigma$. In conclusion, the described procedure is able to reconstruct the statistics of the stochastic quantum entropy production without directly measuring the joint probabilities $p(a^{\text{fin}}_k,a^{\text{in}}_m)$, which instead to realize all the combinatorics concerning the measurement outcomes would require a greater number of measurements, scaling with the square of the values assumed by $\sigma$.

As a final remark, note that this stochastic approach to the quantification of irreversibility can also be applied to systems composed of more than a single particle. Indeed, in such a case the method would rely on considering as effective dimension of the composite system the one that would allow for a unital dynamics. Then, the entropy production of each subsystem would be characterised as a function of the entropy generated by the whole system. 

\section{An alternative formulation to entropy production}
\label{alternative}

This Section aims at pointing out a relevant shortcoming of the formulation of entropy production in terms of the relative entropy, as discussed above. In doing so, we shall highlight a potential resolution of such issues based on the use of generalised entropy functions. The focus of our analysis is that of open quantum systems and, in order to fix the ideas, we shall assume that the dynamics of the system may be modeled by a Lindblad master equation of the form 
\begin{equation}\label{single:master}
\frac{\ud \rho_t}{\ud t} = - i [H, \rho_t] + \mathcal{D}(\rho_t),
\end{equation}
where, as before, $\rho_t$ is the density matrix of the system, $H$ is its Hamiltonian, while ${\cal D}(\rho_t)$ describes the dissipative process arising from its coupling to the external reservoir. In these conditions, it is often convenient to identify the formal contributions to the change of total entropy $\Sigma(t)$ of the state of the system in terms of the equation 
\begin{equation}\label{single:S_sep}
\frac{\ud \Sigma(t)}{\ud t} = \Pi(t)- \Phi(t),
\end{equation}
where $\Pi \geq 0$ is the {entropy production rate} and $\Phi$ is the {entropy flux rate}, from the system to the environment. The entropy production rate $\Pi$ is expected to be non-zero as long as the system is out of equilibrium. This includes transient states and  non-equilibrium steady-states (NESSs), where $\ud \Sigma(t)/\ud t= 0$ and therefore $\Pi = \Phi > 0$. 
The quantities $\Pi$ and $\Phi$ are not direct observables and must therefore be related to experimentally accessible  quantities  via a theoretical framework.

Let $\rho^*$ denote the target state of $\mathcal{D}(\rho_t)$ (for thermal baths $\rho^* = \rho_{\text{eq}} = e^{-\beta H}/Z$). The formulation highlighted in the previous Section of this chapter in terms of the relative entropy would lead us to the following expression for the entropy production rate~\cite{Breuer2003,Breuer2007,Deffner2011}
\begin{equation}\label{Pi:Breuer}
\Pi= - \frac{\ud }{\ud t} S(\rho_t ||  \rho^*).
\end{equation}
While we have already commented on the fact that this formulation satisfies several expected properties for an entropy production, it is worth pointing out that, for a thermal bath,  Eq.~(\ref{Pi:Breuer}) may be factored in the form of Eq.~(\ref{single:S_sep}), with $\Sigma(t)$ being equal to the von Neumann entropy, so that
\begin{equation}\label{Phi_classical}
\Phi(t) = - \frac{1}{T} \tr\bigg[ H \mathcal{D}(\rho)\bigg] \equiv \frac{\Phi_E}{T},
\end{equation}
where $\Phi_E$ denotes the energy flux from the system to the environment. 
This is the well known  Clausius equality of classical thermodynamics.

Despite their clear physical interpretation, a unified approach for the formulation of entropy production beyond the limitations of such educated cases is still lacking: quantum systems  open up the possibility for exploring environmental systems that go beyond the paradigm of equilibrium baths. Striking instances of this are dephasing noises and squeezed baths, whose description extends beyond the usual paradigms of equilibrium baths. Moreover, and quite remarkably, although $\ud \Sigma/\ud t$ remains finite, Eqs.~(\ref{Pi:Breuer}) and (\ref{Phi_classical}) diverge in the limit of zero temperature of the reservoir, owing to the divergence of the relative entropy when the reference state tends to a pure state \cite{Abe2003,Audenaert2013}. This divergence is clearly an inconsistency of the theory. 

Fortunately, a theory of entropy production  that  is applicable to systems exposed to non-equilibrium reservoirs and that cures the divergence at zero temperature is possible~\cite{Jader}.
The key to such alternative theory is the replacement of the von Neumann entropy, which is the pillar upon which the formulations illustrated in the previous Sections have been built, with the R\'enyi-2 entropy. The latter has a similar behaviour to von Neumann's, but is much more convenient to manipulate, and is not pathological when pure reference states are considered. 

In order to provide a framework where analytic expressions are possible, which will serve the purpose of illustration, we focus on bosonic systems characterized by Gaussian states. 
In this case the R\'enyi-2 entropy coincides (up to a constant) with the {Wigner entropy}~\cite{Adesso2012}
\begin{equation}\label{single:S}
S_{\cal W} = - \int \ud^2\alpha\; {\cal W}(\alpha^*,\alpha) \ln {\cal W}(\alpha^*,\alpha),
\end{equation}
where ${\cal W}(\alpha^*,\alpha)$ is the Wigner function and the integral is over the entire complex plane. Gaussian states have a positive Wigner function, which ensures that $S_{\cal W}$ is real.
This link between the R\'enyi-2 entropy and the Wigner entropy allow for a fundamental simplification of the problem, since one may map the  open system dynamics into  a Fokker-Planck equation for ${\cal W}$ and hence employ tools of classical stochastic processes 
to obtain  simple expressions for  $\Pi$ and $\Phi$.
This idea was already used in 
Refs.~\cite{Brunelli2016a,Brunelli2016} via a quantum-to-classical correspondence to treat the case of simple heat baths. Ref.~\cite{Jader} has instead demonstrated the possibility to successfully address squeezed and dephasing reservoirs, while keeping the entropy flux and entropy production rate finite for a system in contact with a thermal reservoir at zero  temperature. Remarkably, for an harmonic oscillator, compact and physically clear expressions are possible for such quantities. Following Ref.~\cite{Jader}, we have
\begin{equation}\label{single:Phi-single:Pi}
\begin{aligned}
\Pi&= \frac{4}{\gamma(\overline{n}+1/2)}  \int\ud^2 \alpha \frac{|J({\cal W})|^2}{{\cal W}},\\
\Phi &= \frac{\gamma}{\overline{n}+1/2} (\langle N \rangle - \overline{n}),
\end{aligned}
\end{equation} 
where $\gamma$ is the damping rate of the oscillator, $\langle N\rangle$ is the mean excitation number for the harmonic oscillator, $\overline{n}$ is the analogous quantity for the bath, and
\begin{equation}
\label{singleJ}
J({\cal W})= \frac{\gamma}{2} \bigg[ \alpha {\cal W} + (\overline{n}+1/2) \partial_{\alpha^*} {\cal W}\bigg],
\end{equation}
with $\alpha$ the complex phase-space variable of the Wigner function. It can be shown that $J$ is in a relation with the phase-space version of the dissipator ${\cal D}$ akin to a continuity equation. Specifically~\cite{Jader}
\begin{equation}
\mathcal{D}({\cal W}) = \partial_\alpha J({\cal W}) + \partial_{\alpha^*} J^*({\cal W}),
\end{equation}
which leads to the interpretation of $J({\cal W})$ as an {\it irreversible} component of the probability current that is null only when the target state is a thermal one (stating the nullity of all probability currents, in this case). The inspection of Eq.~\eqref{single:Phi-single:Pi} shows that, even for zero-temperature baths, $\Phi$ remains finite. In light of the relation between total entropy rate, $\Phi$ and $\Pi$, we deduce that also $\Pi$ does not diverge in such a limit, therefore providing a much more satisfactory result than the one arising from an approach based on the von Neumann entropy.

The generalization to other types of baths is straightforward~\cite{Jader}. 

\section{conclusions}
We have illustrated the current formulation of entropy production in both closed and open systems, highlighting its features and issues, and proposing suitable modifications aimed at defining a full-fledged framework for the characterisation of irreversibility. 

\acknowledgments

TBB acknowledges support from National Research Foundation (Singapore), Ministry of Education (Singapore), and United States Air Force Office of Scientific Research (FA2386-15-1-4082). He acknowledges helpful discussions with R. M. Serra and his group at Universidade Federal do ABC, while developing some of the work reported in this Chapter as part of his Ph.D. thesis. SG gratefully acknowledges Filippo Caruso, Stefano Ruffo, Andrea Trombettoni, and Matthias M. Mueller for useful discussions and an always constructive motivation, especially during the last period of his PhD thesis. GTL would like to acknowledge the S\~ao Paulo Research Foundation, under grant number 2016/08721-7. JPS would like to acknowledge the financial support from the CAPES (PNPD program) for the postdoctoral grant. MP thanks the DfE-SFI Investigator Programme (grant 15/IA/2864), the Royal Society Newton Mobility Grant NI160057, and the H2020 collaborative Project TEQ.

\end{document}